# Quantum Phase Transition in a Modified Jaynes-Cummings Model

Moorad Alexanian

*Department of Physics and Physical Oceanography*
*University of North Carolina Wilmington, Wilmington, NC 28403-5606*

E-mail: alexanian@uncw.edu

**Abstract.** We introduce a modified Jaynes-Cummings model with single-photon cavity radiation field but with the atomic system instead of exchanging a single photon as in the Jaynes-Cummings model, it exchanges instead a squeezed photon with squeezing parameter $r$. This allows us to interpolate between the Rabi model ($r = \infty$) and the Jaynes-Cummings model ($r = 0$) by varying $r$. The model exhibits a quantum phase transition. Accordingly, the quantum phase transition realized in the Rabi model, giving rise to superradiance, also occurs in the Jaynes-Cummings model.

**Keywords:** Rabi model, Jaynes-Cummings model, squeezed state, quantum phase transition, superradiance

## 1. Introduction

The Jaynes-Cummings model of a two level atomic system has been modified recently whereby the single-mode photon radiation field has been replaced by a single-mode squeezed coherent photon radiation field [1]. This modified Jaynes-Cummings model (JCM) was used to study collapse and revival and the behavior was shown to differ considerably from the collapse and revival in the Jaynes-Cummings model. It is interesting that the Rabi model exhibits a normal to superradiance quantum phase transition [2]. The quantum phase transition in the Rabi model has been demonstrated experimentally using a $^{171}$Yb$^+$ ion in a Paul trap [3]. A more recent experiment has been proposed to show the quantum phase transition of the Jaynes-Cummings model by modulating the transition frequency of a two-level system in a quantum Rabi model with strong coupling [4]. The Dicke Hamiltonian, a simple quantum-optical model, exhibits a zero-temperature quantum phase transition. Numerical results have been presented that at this transition the system changes from being quasi-integrable to quantum chaotic [5]. Specifically, the Dicke model with a single two-level system is called the Rabi model.

In this paper, we introduce a different modified Jaynes-Cummings model from that of Ref.1. Here, the radiation field is given by a single-mode photon while the atomic system exchanges squeezed photons instead of single photons. This present model encompasses both the Jaynes-Cummings model as well as the Rabi model as limiting cases and exhibits a quantum phase transition from a normal phase to a superradiance phase. Accordingly, the quantum phase transition present in the Rabi model occurs also in the Jaynes-Cummings model. This paper is arranged as follows. In Sec. 2, the modified Jaynes-Cummings model is introduced. In Sec. 3, a unitary transformation is used to eliminate the coupling terms between spin up and spin down subspaces. In Sec. 4, the Bogoliubov transformation is used to diagonalize the Hamiltonian deduced in Sec. 3 to determine the quantum phase transition. In Sec. 5, the results determining the quantum phase transition is extended to the region where superradiance occurs. Finally, Sec. 6 summarizes our results.

## 2. Modified Jaynes-Cummings model

In a recent paper [1] the creation and annihilation $\hat{B}$ and $\hat{B}^\dagger$ operators, respectively, for the squeezed photons are given by, $\hat{a}$ and $\hat{a}^\dagger$ where are the photon creation and annihilation operators,

$$\hat{B} = \hat{S}(\zeta)\hat{a}\hat{S}(-\zeta) = \cosh(r)\hat{a} + e^{i\varphi}\sinh(r)\hat{a}^\dagger \tag{1}$$

and

$$\hat{B}^\dagger = \hat{S}(\zeta)\hat{a}^\dagger\hat{S}(-\zeta) = e^{-i\varphi}\sinh(r)\hat{a} + \cosh(r)\hat{a}^\dagger, \tag{2}$$

with inverses

$$\hat{a} = \hat{S}(-\zeta)\hat{B}\hat{S}(\zeta) = \cosh(r)\hat{B} - e^{i\varphi}\sinh(r)\hat{B}^\dagger \tag{3}$$

and

$$\hat{a}^\dagger = \hat{S}(-\zeta)\hat{B}^\dagger\hat{S}(\zeta) = -e^{-i\varphi}\sinh(r)\hat{B} + \cosh(r)\hat{B}^\dagger, \tag{4}$$

where

$$\hat{S}(\zeta) = \exp\left(-\frac{\zeta}{2}\hat{a}^{\dagger 2} + \frac{\zeta^*}{2}\hat{a}^2\right) \tag{5}$$

is the squeezing operator with $\zeta = r\exp(i\varphi)$. Note that $[\hat{B}, \hat{B}^\dagger] = 1$ follows from $[\hat{a}, \hat{a}^\dagger] = 1$.

Consider the modified JCM Hamiltonian

$$\hat{H}_{MJC} = \hbar\omega_c \hat{a}^\dagger \hat{a} + \hbar\omega_a \frac{\hat{\sigma}_z}{2} + \frac{\hbar\Omega}{2}(\hat{\sigma}_+ \hat{B} + \hat{\sigma}_- \hat{B}^\dagger), \tag{6}$$

where $\omega_a = \omega_2 - \omega_1$, with $\hbar\omega_1$, $\hbar\omega_2$ are the energies of the uncoupled states /1⟩ and /2⟩, respectively, and $\omega_c$ is the frequency of the field mode. The system can be in two possible states /i⟩, i = 1, 2 with /1⟩ being the ground state of the system and /2⟩ being the excited state, respectively. The transition in the JCM is based on the exchange of one photon, here the transition is via a squeezed photon. The four Paul operators are

$$\begin{aligned} 1 &= |2\rangle\langle 2| + |1\rangle\langle 1| \\ \hat{\sigma}_3 &= |2\rangle\langle 2| - |1\rangle\langle 1| \\ \hat{\sigma}_+ &= |2\rangle\langle 1| = \frac{\hat{\sigma}_x + i\hat{\sigma}_y}{2} \\ \hat{\sigma}_- &= |1\rangle\langle 2| = \frac{\hat{\sigma}_x - i\hat{\sigma}_y}{2}. \end{aligned} \tag{7}$$

It was indicated in Ref.1 that for $r \gg 1$, the Hamiltonian (6) approaches the Rabi model albeit in the ultrastrong coupling limit $\Omega e^r$. Note, however, that (6) reduces to the precise Rabi model provided $\Omega$ vanishes as $e^{-r}$ and so we obtain the exact Rabi model albeit with a finite coupling constant, viz., $\lambda \to -\hbar\Omega e^r/4$ as $r \to \infty$,

$$\hat{H}_{RM} = \hbar\omega_c \hat{a}^\dagger \hat{a} + \hbar\omega_a \frac{\hat{\sigma}_z}{2} - \hbar\lambda(\hat{a} + \hat{a}^\dagger)\hat{\sigma}_x. \tag{8}$$

Similarly, one obtains the Jaynes-Cummings model in the limit $r \to 0$, and so

$$\hat{H}_{JCM} = \hbar\omega_c \hat{a}^\dagger \hat{a} + \hbar\omega_a \frac{\hat{\sigma}_z}{2} + \frac{\hbar\Omega}{2}(\hat{\sigma}_+ \hat{a} + \hat{\sigma}_- \hat{a}^\dagger). \tag{9}$$

Accordingly, the modified Jaynes-Cummings model (6) spans continuously from the Jaynes-Cummings to the Rabi models by varying the squeezing parameter $r$. It is clear that the existence of a quantum phase transition in the Rabi model would imply a quantum phase transition in the modified Jaynes-Cummings model and so also in the Jaynes-Cummings model itself.

Recall from Ref. 1 that (6) can be expressed in terms of photon creation and annihilation operators and so

$$\hat{H}_{MJC} = \hbar\omega_c \hat{a}^\dagger \hat{a} + \hbar\omega_a \frac{\hat{\sigma}_z}{2} + \frac{\hbar\Omega}{2}\left[\cosh(r)(\hat{\sigma}_+\hat{a} + \hat{\sigma}_-\hat{a}^\dagger) + \sinh(r)(\hat{\sigma}_-\hat{a} + \hat{\sigma}_+\hat{a}^\dagger)\right], \tag{10}$$

where we have chosen $\varphi = 0$ and the last term does not occur when making the rotating-wave approximation.

## 3. Unitary transformation

The unitary transformation of (6)
$$\hat{U}^\dagger \hat{H}_{MJC} \hat{U}, \tag{11}$$
where
$$\hat{U} = e^{-v(\sigma_+ \hat{B}^\dagger - \sigma_- \hat{B})} \tag{12}$$

with $v$ real is such that it decouples terms between spin $\mathcal{H}_\downarrow$ $\mathcal{H}_\uparrow$ subspaces and since $\langle \downarrow |\sigma_\pm| \downarrow \rangle = 0$. One has that

$$\hat{H} \equiv \langle \downarrow |\hat{U}^\dagger \hat{H}_{MJC} \hat{U}| \downarrow \rangle = A\hat{a}^\dagger \hat{a} + B + C(\hat{a} + \hat{a}^\dagger)^2, \tag{13}$$

where
$$A = \hbar\omega_c(1 + v^2) + \hbar v(\Omega + \omega_a v)e^{-2r}, \tag{14}$$

$$B = -\frac{\hbar\omega_a}{2}(1 - v^2) + \frac{\hbar v}{2}(\Omega + \omega_a v)e^{-2r} + \hbar\omega_c v^2 \cosh^2(r), \tag{15}$$

and
$$C = -\frac{\hbar v\Omega}{2}\cosh(2r) + \frac{\hbar\omega_a v^2}{2}\sinh(2r). \tag{16}$$

In the above derivation, we have used the leading terms in the Baker–Campbell–Hausdorff formula

$$e^{\hat{X}}\hat{Y}e^{-\hat{X}} = \hat{Y} + [\hat{X},\hat{Y}] + \frac{1}{2!}[\hat{X},[\hat{X},\hat{Y}]] + \cdots \tag{17}$$

## 4. Quantum phase transition

Consider the diagonalization of the Hamiltonian (13)

$$\hat{H} = A\hat{a}^\dagger \hat{a} + B + C(\hat{a} + \hat{a}^\dagger)^2. \tag{18}$$

This is accomplished with the aid of the Bogoliubov transformation of linear boson operators

$$\hat{a} = \cosh(\beta)\hat{b} + \sinh(\beta)\hat{b}^\dagger$$
$$\hat{a}^\dagger = \cosh(\beta)\hat{b}^\dagger + \sinh(\beta)\hat{b}. \tag{19}$$

The cancellation of the terms $\hat{b}\hat{b} + \hat{b}^\dagger\hat{b}^\dagger$ implies that

$$A\cosh(\beta)\sinh(\beta) + C\left[\cosh(\beta) + \sinh(\beta)\right]^2 = 0 \tag{20}$$

and so

$$e^{-4\beta} = \frac{A + 4C}{A}. \tag{21}$$

Hamiltonian (18) becomes

$$\hat{H} = B + A\sinh(\beta)\left[\sinh(\beta) - \cosh(\beta)\right] + A\left[\sinh(\beta) - \cosh(\beta)\right]^2 \hat{b}^\dagger\hat{b}, \tag{22}$$

or

$$\hat{H} = B - \frac{A}{2} + \sqrt{A(A+4C)}\,(\hat{b}^\dagger\hat{b} + 1), \tag{23}$$

where the constants $A$, $B$, $C$ are given by (14), (15), (16), respectively. The quantum phase transition is characterized by $A(A + 4C) = 0$. There are two possible cases, $A = 0$ and/or $A + 4C = 0$.

### A. Case A=0

One obtains for the real variable $v$ associated with the unitary transformation (12) by setting $A = 0$ in (14)

$$v = \frac{-\Omega \pm \sqrt{\Omega^2 - 4\omega_c(\omega_c + \omega_a e^{-2r})e^{4r}}}{2(\omega_a + \omega_c e^{2r})}. \tag{24}$$

In order to remain in one quantum phase and thus avoid a bifurcation of the value of $v$, which suggests a different quantum phase for $\Omega^2 > \Omega_c^2$, we must have

$$\Omega_c^2 = 4\omega_c(\omega_c + \omega_a e^{-2r})e^{4r}. \tag{25}$$

The critical point for the Jaynes-Cummings model ($r = 0$) is then

$$\Omega_{MJC}^2 = 4\omega_c(\omega_a + \omega_c). \tag{26}$$

However, for the Rabi model ($\hbar\Omega \to -4\lambda e^{-r}$ as $r \to \infty$) (25) implies

$$-\infty < \lambda < \infty, \tag{27}$$

which gives no specific value for the critical point for the Rabi model.

### B. Case A+4C=0

One obtains for the real variable $v$ associated with the unitary transformation (12) by setting $A+4C = 0$ and using (14) and (16)

$$v = \frac{\Omega \pm \sqrt{\Omega^2 - 4\omega_c(\omega_a + \omega_c e^{-2r})e^{-2r}}}{2(\omega_a + \omega_c e^{-2r})}. \tag{28}$$

As in the previous case for $A = 0$, we must have a critical point in order to remain in one quantum phase and so

$$\Omega_c^2 = 4\omega_c(\omega_a + \omega_c e^{-2r})e^{-2r}, \tag{29}$$

where a differing quantum phase occurs for $\Omega^2 > \Omega_c^2$. The critical point for the Jaynes-Cummings model ($r = 0$) is then

$$\Omega_{MJC}^2 = 4\omega_c(\omega_a + \omega_c). \tag{30}$$

which agrees with the previous for $A = 0$ given by (26).
However, for the Rabi model ($\hbar\Omega \to -4\lambda e^{-r}$ as $r \to \infty$) (29) implies

$$4\lambda_c^2 = \omega_a \omega_c \hbar^2. \tag{31}$$

with critical point

$$\lambda_{MB} = \pm\frac{\hbar}{2}\sqrt{\omega_a \omega_c}. \tag{32}$$

The variables that appear in the diagonalized Hamiltonian $\hat{H}$ (23) are as follows

$$\frac{1}{2}(2B - A) = -\frac{\hbar}{2}(\omega_a + \omega_c) + \frac{\hbar\Omega^2}{8}\frac{\omega_a + \omega_c \cosh(2r)}{(\omega_a + \omega_c e^{-2r})^2}, \tag{33}$$

$$A = \hbar\omega_c + \frac{\hbar\Omega^2}{4(\omega_a + \omega_c e^{-2r})^2}\left[\omega_c(1 + 2e^{-4r}) + 3\omega_a e^{-2r}\right], \tag{34}$$

and

$$A + 4C = \hbar\omega_c - \frac{\hbar\Omega^2 e^{2r}}{4(\omega_a + \omega_c e^{-2r})}, \tag{35}$$

where we have chosen $v = \Omega/[2(\omega_a + \omega_c e^{-2r})]$ to evaluate A, B, and C in (14)-(16).
One obtains from the modified Jaynes-Cummings model (23) for $r = 0$ with the aid of (33)-(35)

$$\hat{H}_{JCM} = -\hbar(\omega_a + \omega_c)/2 + \frac{\hbar\Omega^2}{8(\omega_a + \omega_c)} + \hbar\omega_c\sqrt{\left(1 + \frac{3\Omega^2}{4\omega_c(\omega_a + \omega_c)}\right)\left(1 - \frac{\Omega^2}{4\omega_c(\omega_a + \omega_c)}\right)} \, (\hat{b}^\dagger\hat{b} + 1). \tag{36}$$

Similarly, one obtains from the modified Jaynes-Cummings model (23) for $r = \infty$ with the aid of (33)-(35) and since $\hbar\Omega \to -4\lambda e^{-r}$

$$\hat{H}_{BM} = -\hbar(\omega_a + \omega_c)/2 + \frac{\omega_c \lambda^2}{\hbar\omega_a^2} + \hbar\omega_c\sqrt{1 - \frac{4\lambda^2}{\hbar^2 \omega_a \omega_c}} \, (\hat{b}^\dagger\hat{b} + 1). \tag{37}$$

Recall that in Ref.2, the authors consider $\omega_c/\omega_a \to 0$ and so

$$\hat{H}_{BM} \to -\hbar\omega_a/2 + \hbar\omega_c\sqrt{1 - \frac{4\lambda^2}{\hbar^2\omega_a\omega_c}}\ \hat{b}^\dagger\hat{b}, \tag{38}$$

as $\omega_c/\omega_a \to 0$ in agreement with their result.

## 5. Superradiance

The existence of a different quantum phase for $\Omega^2 > 4\omega_c(\omega_a + \omega_c e^{-2r})e^{-2r}$ is associated with the bifurcation of the value of $v$ in (28). In order to obtain results for this new quantum state, we consider a Glauber displacement operator on the modified Jaynes-Cummings Hamiltonian (10)

$$\hat{H}_d(\alpha) = \hat{D}^\dagger(\alpha)\hat{H}_{MJC}\hat{D}(\alpha) = \hbar\omega_c(\hat{a}^\dagger + \alpha)(\hat{a} + \alpha) + \hbar\omega_a\frac{\hat{\sigma}_z}{2} + \frac{\hbar\Omega\alpha e^r}{2}(\hat{\sigma}_+ + \hat{\sigma}_-)$$
$$+ \frac{\hbar\Omega}{2}\left[\cosh(r)(\hat{\sigma}_+\hat{a} + \hat{\sigma}_-\hat{a}^\dagger) + \sinh(r)(\hat{\sigma}_-\hat{a} + \hat{\sigma}_+\hat{a}^\dagger)\right] \tag{39}$$

where

$$\hat{D}(\alpha) = e^{\alpha(\hat{a}^\dagger - \hat{a})} \tag{40}$$

and $\alpha$ real. The eigenstates of the atomic part of the Hamiltonian (39), viz., $\hbar\omega_a\hat{\sigma}_z/2 + \hbar\alpha e^r\Omega\hat{\sigma}_x/2$, are

$$|\tilde{\uparrow}\rangle = \cos(\theta)|\uparrow\rangle + \sin(\theta)|\downarrow\rangle, \qquad |\tilde{\downarrow}\rangle = -\sin(\theta)|\uparrow\rangle + \cos(\theta)|\downarrow\rangle \tag{41}$$

with $\tan(2\theta) = \alpha\Omega e^r/\omega_a$. The new transition frequency is given $\tilde{\Omega} = \sqrt{\omega_a^2 + \alpha^2\Omega^2 e^{2r}}$, by $\pm\hbar\tilde{\Omega}/2$ where are the eigenvalues of the atomic part of the Hamiltonian (39). The relation between Pauli matrices is as follows

$$\hat{\sigma}_x = \cos(2\theta)\hat{\tau}_x + \sin(2\theta)\hat{\tau}_z$$
$$\hat{\sigma}_y = \hat{\tau}_y \tag{42}$$
$$\hat{\sigma}_z = -\sin(2\theta)\hat{\tau}_x + \cos(2\theta)\hat{\tau}_z.$$

In terms of Pauli matrices $\hat{\tau}_{x,y,z}$ in the $|\tilde{\uparrow}\rangle, |\tilde{\downarrow}\rangle$ basis, (39) becomes

$$\tilde{H}_{MJC}(\alpha) = \hbar\omega_c\hat{a}^\dagger\hat{a} + \hbar\omega_c\alpha^2 + \frac{\hbar\Omega^2 e^{2r}}{8\omega_c}\hat{\tau}_z + \frac{\hbar\Omega\cos(2\theta)e^r}{4}(\hat{a} + \hat{a}^\dagger)\hat{\tau}_x. \tag{43}$$

The Hamiltonian (43) is of the generic Rabi form

$$\hat{H}_g = J\hat{a}^\dagger\hat{a} + K + L\hat{\tau}_z + M(\hat{a} + \hat{a}^\dagger)\hat{\tau}_x, \tag{44}$$

where $J, K, L, M$ are real constants, that can be diagonalized with the aid of the unitary $\hat{S}(\mu) = \exp[\mu(\hat{a} + \hat{a}^\dagger)(\hat{\sigma}_+ - \hat{\sigma}_-)]$ transformation with real $\mu$ and so

$$\hat{S}^\dagger(\mu)\hat{H}_g\hat{S}(\mu) = J\hat{a}^\dagger\hat{a} + K + L\hat{\tau}_z - 2\mu(M + \mu L)(\hat{a} + \hat{a}^\dagger)^2\hat{\tau}_z. \tag{45}$$

Result (45) follows by keeping the lowest order terms in the Baker–Campbell–Hausdorff formula (17) which upon projection onto $\mathcal{H}_{\downarrow}$ becomes

$$\hat{H}_{\downarrow} = J\hat{a}^{\dagger}\hat{a} + K - L + 2\mu(M + \mu L)(\hat{a} + \hat{a}^{\dagger})^2. \quad (46)$$

The excitation energy follows from (23) and so

$$\epsilon = J\sqrt{1 + 8\mu(M + \mu L)/J}. \quad (47)$$

Now

$$\mu = \frac{\tilde{\lambda}}{\tilde{\Omega}} = \frac{-M}{2L} = -\frac{\omega_c \cos(2\theta)}{\Omega e^r}, \quad (48)$$

where $-\tilde{\lambda}$ is the coefficient of the $(\hat{a} + \hat{a}^{\dagger})\hat{\tau}_x$ term in (43) and $\tilde{\Omega}/2$ is the coefficient of the $\tau_z$ term in (43) [2] that yields

$$\epsilon = J\sqrt{1 - \cos^2(2\theta)}, \quad (49)$$

where

$$\cos(2\theta) = \frac{\omega_a}{\sqrt{\omega_a^2 + \alpha^2\Omega^2 e^{2r}}}. \quad (50)$$

In what follows, we determine the expression for $\cos^2(2\theta)$ for both the Rabi model, as well as, the Jaynes-Cummings model by choosing the value of $\alpha^2$.

**A. Rabi model**

One obtains the Rabi model in the limit $\hbar\Omega \to -4\lambda e^r$ as $r \to \infty$ thus (50) yields

$$\cos(2\theta) = \frac{\omega_a}{\sqrt{\omega_a^2 + 16\lambda^2\alpha^2/\hbar^2}}. \quad (51)$$

In Ref.2, the values of $\alpha$ and $\lambda$ are chosen as follows, $\alpha^2 = (\omega_a/4g^2\omega_c)(g^4 - 1)$ and $\lambda^2 = \omega_a\omega_c\hbar^2 g^2/4$ and one obtains for the excitation energy

$$\epsilon = \omega_c\hbar\sqrt{1 - g^{-4}} \quad (52)$$

with critical point value $g_c = 1$.

**B. Jaynes-Cummings model**

One obtains the Jaynes-Cummings model for $r = 0$ thus, with the aid of (50)

$$\cos(2\theta) = \frac{\omega_a}{\sqrt{\omega_a^2 + \alpha^2\Omega^2}}. \quad (53)$$

We chose

$$\alpha^2 = (\omega_a^2/\Omega^2)(\Omega^4/\Omega_{JMC}^4 - 1) \quad (54)$$

and one obtains for the excitation energy

$$\epsilon = \omega_c \hbar \sqrt{1 - \Omega_{JMC}^4/\Omega^4}. \qquad (55)$$

with critical point value $\Omega_c = \Omega_{JMC}$.

## 6. Conclusions

A modified Jaynes-Cummings model whereby the atomic transition is associated with squeezed photons rather than ordinary photons interpolates between the Jaynes-Cummings model and the Rabi model by varying the squeezing parameter. Accordingly, a quantum phase transition in the modified Jaynes-Cummings model that is in accord with the quantum phase transition to superradiance in the Rabi model will be associated with a similar quantum phase transition to superradiance in the Janyes-Cummings model.